\documentclass[aps,prl,twocolumn,letterpaper]{revtex4}
 
\usepackage{amsmath,amssymb,amsfonts} \usepackage{graphicx}
 
\begin{document}
 
\title{Effect of epitaxial strain on the spontaneous polarization of
  thin film ferroelectrics}
 
\date{\today}
 
\author{Claude Ederer}
\affiliation{Materials Research Laboratory and Materials Department,
  University of California, Santa Barbara, CA 93106, U.S.A.}
\email{ederer@mrl.ucsb.edu}
\author{Nicola A.~Spaldin}
\affiliation{Materials Research Laboratory and Materials Department,
  University of California, Santa Barbara, CA 93106, U.S.A.}

\begin{abstract}
We investigate the variation of the spontaneous ferroelectric
polarization with epitaxial strain for BaTiO$_3$, PbTiO$_3$,
LiNbO$_3$, and BiFeO$_3$ using first principles calculations. We find
that while the strain dependence of the polarization is very strong in
the simple perovskite systems BaTiO$_3$ and PbTiO$_3$ it is only weak
in LiNbO$_3$ and BiFeO$_3$. We show that this different behavior can
be understood purely in terms of the piezoelectric and elastic
constants of the unstrained bulk material, and we discuss several
factors that determine the strain behavior of a certain material.
\end{abstract}

\maketitle

The possible application of ferroelectric materials in microelectronic
devices has led to strong interest in the properties of thin film
ferroelectrics \cite{Dawber/Rabe/Scott:2005}. One important question
in this context is how epitaxial strain, which is incorporated in the
ferroelectric material due to the lattice mismatch with the substrate,
affects the ferroelectric characteristics of the thin film. It has
been demonstrated that epitaxial strain can have drastic effects, such
as inducing ferroelectricity at room temperature in otherwise
paraelectric SrTiO$_3$ \cite{Haeni_et_al:2004}, or increasing the
ferroelectric Curie temperature of BaTiO$_3$ by nearly 500$^\circ$C
and the remanent polarization by 250~\% compared with the
corresponding bulk values \cite{Choi_et_al:2004}. Based on these
observations it is often assumed that the strong sensitivity to
epitaxial strain is a common feature of all ferroelectrics. 

Indeed, this assumption has been supported by first principles
calculations for several simple $AB$O$_3$ perovskite ferroelectrics
\cite{Neaton/Hsueh/Rabe:2002,Neaton/Rabe:2003,Bungaro/Rabe:2004,Dieguez/Rabe/Vanderbilt:2005}.
Strain is introduced in these calculations by fixing the lattice
constants corresponding to the lateral directions of the substrate
while relaxing all remaining structural parameters, according to the
minimum of the total energy under the epitaxial constraint. This makes
it possible to isolate the pure strain effect from other effects
present in real thin film samples such as structural defects, chemical
inhomogeneities, and interface effects. For example, in
Refs.~\onlinecite{Bungaro/Rabe:2004} and
\onlinecite{Dieguez/Rabe/Vanderbilt:2005} the change in phase
stability due to epitaxial strain was investigated by first principles
techniques for several simple perovskite systems including the
prototype ferroelectrics BaTiO$_3$ and PbTiO$_3$. It was shown that
for these systems a series of consecutive phase transitions occur
which effectively rotate the polarization from ``out-of-plane'' for
compressive epitaxial strain to ``in-plane'' for tensile epitaxial
strain. Also, the magnitude of the spontaneous polarization within the
different phases was shown to be strongly strain dependent. In
contrast, we have recently shown that in BiFeO$_3$, a multiferroic
system where ferroelectricity coexists with magnetic order, the
magnitude of the polarization barely changes even for relatively large
strain values of $\pm$3~\% \cite{Ederer/Spaldin_2:2005}.

In the present work we investigate the factors that lead to a strong
strain dependence of the spontaneous polarization in some materials
and to a relative inertness to epitaxial strain in other materials. To
do this we compare the spontaneous polarization as a function of
epitaxial strain (in a fixed phase) for a variety of different
perovskite-derived ferroelectrics with different structural symmetries
and different mechanisms causing the ferroelectric displacements. We
show that the sensitivity to epitaxial strain varies considerably for
different materials and that strong strain dependence is not a
universal feature of all ferroelectrics. Furthermore, we show that the
strain dependence of the polarization for experimentally relevant
strain values in all systems can be understood in terms of the
piezoelectric and elastic constants of the unstrained materials and we
discuss the factors determining the sensitivity to epitaxial strain of
a certain material.

In general, the total change of the spontaneous polarization $P$ to
linear order in the strain components $\epsilon_i$ is given by the
improper piezoelectric tensor, $c_{\alpha i}$ \cite{Martin:1972}:
\begin{equation}
\label{piezo}
\frac{\partial P_\alpha}{\partial \epsilon_i} = c_{\alpha i} \quad .
\end{equation}
Here, $\alpha = 1,2,3$ stands for the usual cartesian components $x$,
$y$, and $z$, whereas $i=1,\dots,6$ denotes the components of the
strain tensor in Voigt notation (see
e.g. Ref.~\onlinecite{Lines/Glass:Book}). For epitaxial strain,
$\epsilon_4 = \epsilon_5 = \epsilon_6 = 0$ and $\epsilon_1 =
\epsilon_2 \neq 0$. Furthermore, the ratio between in-plane and
out-of-plane strain is given by the \emph{Poisson ratio}
$n=-\epsilon_1/\epsilon_3$. In the following we only consider the
situation where the spontaneous polarization is directed perpendicular
to the orientation of the substrate plane, and we use this as the $z$
direction of our coordinate system. The substrate surface is then
oriented parallel to the $x$-$y$ plane. In this case, the change in
polarization (to linear order in the strain) is given by:
\begin{equation}
\label{ceff}
\Delta P_3 = \left( 2 c_{31} - \frac{c_{33}}{n} \right) \epsilon_1 =
c_\text{eff} \, \epsilon_1 \quad .
\end{equation}
Here we have further assumed that the symmetry is such that $c_{31}$ =
$c_{32}$.

We now investigate the validity of Eq.~\ref{ceff} for experimentally
relevant strain values by comparing directly calculated values of the
spontaneous polarization as a function of epitaxial strain with values
calculated using Eq.~\ref{ceff} and the piezoelectric constants and
Poisson ratios of the corresponding unstrained structure. We also
investigate how the effective piezoelectric constant $c_\text{eff}$,
which describes the change in polarization due to epitaxial strain,
varies for different materials.

In Ref.~\onlinecite{Ederer/Spaldin_2:2005} several possible
explanations for the different strain behavior in BiFeO$_3$ compared
to systems like BaTiO$_3$/PbTiO$_3$ were suggested: (i) different
structural symmetry: BiFeO$_3$ crystallizes in the rhombohedral $R3c$
structure with a doubled unit cell compared to the ideal perovskite
structure and shows rotated oxygen octahedra in addition to the polar
displacements (see Ref.~\onlinecite{Neaton_et_al:2005}), whereas the
relevant phases \footnote{BaTiO$_3$ is tetragonal in the temperature
range 280~K $\alt$ $T$ $\alt$ 400~K \cite{LB:36}.} for
BaTiO$_3$/PbTiO$_3$ are simple tetragonally distorted perovskite
structures (space group $P4mm$)
\cite{Bungaro/Rabe:2004,Dieguez/Rabe/Vanderbilt:2005}; (ii) different
displacement mechanisms: in BaTiO$_3$/PbTiO$_3$ the $d^0$
configuration of the Ti cation plays a crucial role for the
ferroelectric instability, whereas in the multiferroic system
BiFeO$_3$ the ferroelectricity is solely caused by the Bi 6$s$ lone
electron pair \cite{Hill:2002}; (iii) a general high stability of the
ferroelectric state in BiFeO$_3$, indicated by rather large ionic
displacements and a high ferroelectric Curie temperature of 1123~K
(BaTiO$_3$: 400~K, PbTiO$_3$: 763~K) \cite{LB:36}.

In order to systematically address the possible explanations (i) to
(iii), we study the following systems: the prototypical ferroelectrics
BaTiO$_3$ and PbTiO$_3$; the multiferroic system BiFeO$_3$, both in
its ground state $R3c$ structure and in a hypothetical metastable
$P4mm$ structure, i.e. isostructural to BaTiO$_3$/PbTiO$_3$;
LiNbO$_3$, which is isostructural to BiFeO$_3$, with large ionic
displacements and a very high Curie temperature of 1480~K
\cite{LB:36}, but where, in contrast to the multiferroic system
BiFeO$_3$, the transition metal cation plays an active role in the
ferroelectric instability, similar to the case of BaTiO$_3$/PbTiO$_3$
\cite{Inbar/Cohen:1996,Veithen/Ghosez:2002}.

Values for the spontaneous polarization obtained in this work are
calculated using the Berry phase approach for determining the
electronic contribution to the polarization
\cite{King-Smith/Vanderbilt:1993,Vanderbilt/King-Smith:1994}. The
piezoelectric constants are obtained using the Berry phase approach
and a finite difference method analogous to
Ref.~\onlinecite{Saghi-Szabo/Cohen/Krakauer:1998}. The Poisson ratios
are estimated from the change in the out-of-plane lattice constant
when fixing the in-plane lattice constant and relaxing all ionic
positions. For all calculations we use the \emph{Vienna Ab-initio
Simulation Package} (VASP) \cite{Kresse/Furthmueller_CMS:1996}, which
employs the Projector Augmented Wave (PAW) method within density
functional theory \cite{Bloechl:1994,Kresse/Joubert:1999}. Except
where otherwise noted we use the local spin-density approximation
(LSDA), see Ref.~\onlinecite{Jones/Gunnarsson:1989}. All calculational
parameters are tested to result in good convergence of all quantities
under consideration.

\begin{table}
\caption{Structural parameters calculated for LiNbO$_3$. $a$ and $c$
  are the lattice parameters of the corresponding hexagonal unit
  cell. $z$, $u$, $v$, $w$ are internal structural parameters defined
  in Ref.~\onlinecite{Veithen/Ghosez:2002}.}
\label{table:LNO}
\begin{ruledtabular}
\begin{tabular}{cccccc}
$a$ [\AA] & $c$ [\AA] & $z$ & $u$ & $v$ & $w$ \\
\hline
5.073 & 13.702 & 0.0337 & 0.01257 & 0.0426 & 0.0178
\end{tabular}
\end{ruledtabular}
\end{table}

\begin{table}
\caption{Structural parameters calculated for $P4mm$ BaTiO$_3$,
  PbTiO$_3$, and BiFeO$_3$. $a$ and $c/a$ describe the tetragonal unit
  cell, $\Delta_i$ is the displacement of ion $i$ along the $c$ axis
  (in units of $c$). The notation is the same as used in
  Ref.~\onlinecite{Neaton/Hsueh/Rabe:2002}.}
\label{table:BFO}
\begin{ruledtabular}
\begin{tabular}{lccccc}
 & $a$ [\AA] & $c/a$ & $\Delta_B$ & $\Delta_\text{O$_I$}$ &
 $\Delta_\text{O$_{II}$}$ \\ 
\hline 
BaTiO$_3$ & 3.945 & 1.011 & 0.012 & -0.019 & -0.013 \\
PbTiO$_3$ & 3.870 & 1.040 & 0.034 &  0.087 &  0.101 \\
BiFeO$_3$ & 3.665 & 1.270 & 0.064 &  0.170 &  0.203 \\ 
\end{tabular}
\end{ruledtabular}
\end{table}

The structural parameters calculated in this work corresponding to the
unstrained systems are listed in Table~\ref{table:LNO} for LiNbO$_3$
and in Table~\ref{table:BFO} for the tetragonal systems. All
structural parameters are in good agreement with previous
calculations, see Refs.~\onlinecite{Inbar/Cohen:1996} and
\onlinecite{Veithen/Ghosez:2002} for LiNbO$_3$,
Ref.~\onlinecite{Neaton/Hsueh/Rabe:2002} for BaTiO$_3$, and
Ref.~\onlinecite{Garcia/Vanderbilt:1996} for PbTiO$_3$. For $R3c$
BiFeO$_3$ the structural parameters are listed in Table II of
Ref.~\onlinecite{Neaton_et_al:2005} ($U_\text{eff}$ = 0). All systems
except $P4mm$ BiFeO$_3$ are fully relaxed using the LSDA. For $P4mm$
BiFeO$_3$ calculations are performed using the LSDA+$U$ method
\cite{Dudarev_et_al:1998} with a small $U_\text{eff}$ = 2~eV in order
to ensure the insulating character. The Poisson ratios and the
spontaneous polarization as a function of epitaxial strain were
extracted from Ref.~\onlinecite{Neaton/Hsueh/Rabe:2002} for BaTiO$_3$,
and from Ref.~\onlinecite{Bungaro/Rabe:2004} for PbTiO$_3$. All other
quantities are calculated in this work.

Since BiFeO$_3$ in the metastable $P4mm$ structure has not been
discussed in the literature before, we give a few more details
obtained by our structural relaxation of this system. The energy
minimum of the relaxed $P4mm$ structure of BiFeO$_3$ is
$0.26$~eV/(formula unit) higher than for the ground state $R3c$
structure (calculated with the same $U_\text{eff}$). The relaxed
$P4mm$ structure is strongly distorted compared to the ideal
perovskite structure, with a $c/a$ ratio of 1.27 and a very large
polarization of 151~$\mu$C/cm$^2$. These values are similar to those
found for the recently predicted isostructural ferroelectric BiGaO$_3$
\cite{Baettig_et_al:2005}.

Before presenting our main results, we clarify two technical
points. First, since the polarization of a periodic solid is only well
defined modulo $eR/V$, where $e$ is the electronic charge, $R$ is a
lattice vector, and $V$ is the unit cell volume, the calculation of
the polarization for a certain bulk structure leads to a lattice of
values \cite{King-Smith/Vanderbilt:1993,Vanderbilt/King-Smith:1994}.
The spontaneous polarization $P$ can be obtained by monitoring the
change in polarization $\Delta P$ of an arbitrary chosen ``branch'' of
the polarization lattice, when the structure is deformed from the
original structure into a state with inverted polarization. It has
been shown by King-Smith and Vanderbilt
\cite{King-Smith/Vanderbilt:1993,Vanderbilt/King-Smith:1994} that this
change is independent of the chosen ``path'' leading from the initial
to the final state, as long as the system stays insulating for all
intermediate states along the path. The spontaneous polarization is
then given by $P=\Delta P/2$ and is also independent of the chosen
branch.

\begin{figure}
\includegraphics[width=0.6\columnwidth]{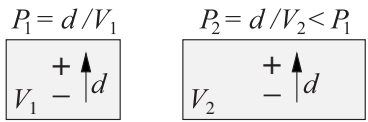}
\caption{Reason for the difference between proper and improper
  piezoelectric response: the polarization $P_2$ is smaller than $P_1$
  although the dipole moment $d$ is unchanged.}
\label{fig:prop}
\end{figure}

Second, it has been pointed out in the literature that the improper
piezoelectric response depends on the specific branch used to
calculate this quantity \cite{Vanderbilt:2000}. The distinction
between \emph{proper} and \emph{improper} piezoelectric response is a
consequence of the definition of the polarization as the dipole moment
per volume, which causes the polarization to vary as a function of
strain even when the dipole moment does not change (see
Fig.~\ref{fig:prop}) \cite{Martin:1972,Vanderbilt:2000}. This pure
volume effect is included in the improper but not in the proper
piezoelectric response. However, the two are trivially related by the
spontaneous polarization (in the case of $c_{33}$ the proper is equal
to the improper piezoelectric response, for $c_{31}$ the proper
piezoelectric response is given by the improper value plus the
spontaneous polarization) \cite{Martin:1972,Vanderbilt:2000}. Here we
report the improper piezoelectric response, which is used in
Eqs.~(\ref{piezo}) and (\ref{ceff}), since they refer to the total
change in spontaneous polarization. We point out that the improper
piezoelectric constant corresponding to the \emph{spontaneous}
polarization is well defined and does not depend on the specific
branch used in the evaluation of this quantity.

\begin{table}
\caption{Improper Piezoelectric constants $c_{33}$, $c_{31}$, and
  $c_\text{eff}$ (in $\mu$C/cm$^2$) as well as the Poisson ratio $n$
  for the various systems. References are given for values not
  calculated in this work.}
\label{table:piezo}
\begin{ruledtabular}
\begin{tabular}{lcccc}
 & $c_{33}$ & $c_{31}$ & $n$ & $c_\text{eff}$ \\ 
\hline 
BaTiO$_3$ & 670 &  30 & 0.65 \cite{Neaton/Hsueh/Rabe:2002} & -971 \\
PbTiO$_3$ & 586 & 103 & 0.58 \cite{Bungaro/Rabe:2004} & -804 \\ 
BiFeO$_3$ ($R3c$)  & 213 &  50 & 1.15 &  -85 \\ 
LiNbO$_3$          & 124 & -75 & 1.5 & -233 \\ 
BiFeO$_3$ ($P4mm$) & 203 & -15 & 0.65 & -342 \\
\end{tabular}
\end{ruledtabular}
\end{table}

The improper piezoelectric constants and Poisson ratios for the five
systems are given in Table~\ref{table:piezo}. Note that for the
tetragonal systems the fourfold axis is used as the $z$ direction of
our coordinate system, whereas for the rhombohedral systems the
threefold axis is used. It is evident that the effective piezoelectric
constants $c_\text{eff}$ describing the effect of epitaxial strain (to
linear order) on the spontaneous polarization are much larger for
BaTiO$_3$ and PbTiO$_3$ than for the other three systems. This is
mainly due to the larger $c_{33}$ in BaTiO$_3$/PbTiO$_3$ but also due
to the smaller Poisson ratio of the $P4mm$ structures \footnote{The
piezoelectric constants for PbTiO$_3$ have also been calculated in
Ref.~\onlinecite{Saghi-Szabo/Cohen/Krakauer:1998} using the
generalized gradient approximation (GGA) instead of the LSDA. It seems
that the LSDA values are slightly larger than the GGA values, but the
difference could also (at least partly) be due to the different
structural parameters used in the two calculations.}. The Poisson
ratios are comparable for systems with the same structural
symmetry. In $P4mm$ BiFeO$_3$ the smaller Poisson ratio leads to a
slightly increased $c_\text{eff}$ compared to $R3c$ BiFeO$_3$ although
the piezoelectric constant $c_{33}$ is similar for both
symmetries. Another observation is that, although $c_{33}$ is smaller
in $R3c$ BiFeO$_3$ than in LiNbO$_3$, $c_\text{eff}$ is larger due to
the different sign of $c_{31}$.

\begin{figure}
\includegraphics*[width=0.4\textwidth]{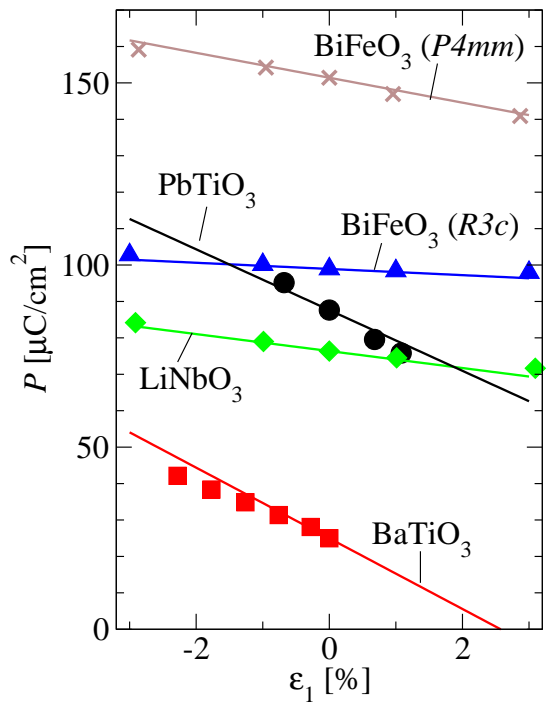}
\caption{(Color online) Spontaneous polarization $P$ for several
  ferroelectrics as a function of epitaxial strain
  $\epsilon_1$. Symbols represent directly calculated values, lines
  are calculated using Eq.~(\ref{ceff}) and the effective
  piezoelectric constants from Table~\ref{table:piezo}. Directly
  calculated values for BaTiO$_3$ and PbTiO$_3$ are taken from
  Refs.~\onlinecite{Neaton/Hsueh/Rabe:2002} and
  \onlinecite{Bungaro/Rabe:2004}, respectively.}
\label{fig}
\end{figure}

Fig.~\ref{fig} compares the directly calculated change in spontaneous
polarization caused by epitaxial strain with the corresponding change
calculated using $c_\text{eff}$ from Table~\ref{table:piezo} for all
five systems. One can see that the linear approximation of
Eq.~\ref{ceff} works well for all systems up to strains of $\pm$
3~\%. Even in the case of BaTiO$_3$ and PbTiO$_3$, where we have used
data from two different sources, the agreement between the
polarization calculated using Eq.~\ref{ceff} and the directly
calculated values is remarkably good. For the systems where all values
are calculated in the present work the agreement is extremely good.

The present study shows that the polarization response to epitaxial
strain for experimentally relevant strain values can be described in
terms of the piezoelectric and elastic constants of the unstrained
system. Two further observations can be made. First, the appearance of
two different terms in Eq.~\ref{ceff} with potentially opposite signs
makes it possible that a system can in principle have rather large
piezoelectric constants, but that due to partial cancellation
$c_\text{eff}$ is rather small. Second, a large value of $c_{33}$ can
eventually be suppressed by a large Poisson ratio. Otherwise, if no
partial cancellation occurs, a system with a large piezoelectric
response will in general also show a strong response of the
spontaneous polarization to epitaxial strain.

The question of whether epitaxial strain enhances the polarization in
ferroelectric thin films therefore leads to the more general problem
of what determines the piezoelectric (and elastic) response of a
certain material. An overview of recent first principles research on
piezoelectricity is given in Ref.~\cite{Bellaiche:2002}. Some insight
can be gained by separating the piezoelectric response into a
``clamped ion'' part and an ``internal strain'' part. The first
measures the strain-induced change in polarization for fixed internal
ionic coordinates whereas the latter gives the contribution resulting
from the relaxation of the ions. It can be shown that the internal
strain part of the piezoelectric tensor contains the inverse of the
force constant matrix \cite{Wu/Vanderbilt/Hamann:2005}. Therefore, if
the force constant matrix has rather small eigenvalues, i.e. low
energy phonon modes, and these modes couple to a certain strain
component, the piezoelectric response can be very large and is then
dominated by the internal strain part. In a soft mode system at least
one phonon energy goes to zero at the ferroelectric transition
temperature and the piezoelectric response diverges
\cite{Lines/Glass:Book}. Far away from the transition temperature the
energy of this soft phonon mode increases again. This can in principle
explain the decrease of the (zero temperature) value of $c_{33}$ in
the series BaTiO$_3$, PbTiO$_3$, BiFeO$_3$, and LiNbO$_3$, i.e. for
increasing Curie temperatures. For $c_{31}$ both contributions to the
piezoelectric constant are comparable and of opposite sign, leading to
no clear trend in magnitude or sign for the different materials.

In summary, regarding the three possible explanations (i) to (iii)
given in Ref.~\onlinecite{Ederer/Spaldin_2:2005} for the different
strain behavior in different systems, we can now conclude that the
structural symmetry (i) mainly affects the value of the Poisson ratio,
whereas the ``general stability'' of the ferroelectric state (iii)
determines the order of magnitude of $c_{33}$. No clear influence of
the different displacement mechanisms (ii) or of the
``multiferroicity'' of BiFeO$_3$ can be identified from our results,
although the displacement mechanism can of course influence the
stability of the ferroelectric state. It seems that in most cases a
large $c_{33}$ leads to a strong dependence of the polarization on
epitaxial strain, and conversely a strong epitaxial strain dependence
is usually connected with a large $c_{33}$. Nevertheless, a small
(large) Poisson ratio can enhance (suppress) this effect and the same
is true for the sign and magnitude of $c_{31}$. It also becomes clear
from our results that not every ferroelectric necessarily shows a
strong sensitivity to epitaxial strain, and that the response for a
certain material can be calculated from its piezoelectric tensor and
Poisson ratio using Eq.~\ref{ceff} for all experimentally accessible
epitaxial strains.

\begin{acknowledgments}
The authors thank Priya Gopal for valuable discussions. This work was
supported by the MRSEC program of the National Science Foundation
under Award No. DMR00-80034.
\end{acknowledgments}

\bibliography{../../literature.bib}

\end{document}